\documentstyle{lamuphys} 
\makeatletter
\let\chapter\hid@chapter
\makeatother
\begin{document}
\def\shalf{\hbox{${\textstyle{\frac{1}{2}}}$}}
\def\quarter{\hbox{${\textstyle{\frac{1}{4}}}$}}
\def\eights{\hbox{${\textstyle{\frac{1}{8}}}$}}
\def\reals{\hbox{\bf R}}
\def\integers{\hbox{\bf Z}}
\newfont{\operator}{cmssdc10}
\def\S{\hbox{\operator S}}
\def\N{\hbox{\operator N}}
\def\n{\hbox{$n^{\flat}$}}
\def\a{\hbox{$a^{\flat}$}}
\def\R3{\hbox{$R^{(3)}$}}
\def\Deltatwo{\hbox{$\Delta^{(2)}$}}
\def\t#1{\hbox{{$\tilde{#1}$}}}
\def\f{\rm f}

\pagenumbering{arabic}
\titlerunning{Time-Symmetric Initial Data}
\title{On the Construction of Time-Symmetric\\
       Black Hole Initial Data\footnote{
       To appear in {\it Black Holes: Theory and Observation},
       edited by F.~Hehl, C.~Kiefer and R.~Metzler, 
       Springer Verlag 1998.}}

\author{Domenico Giulini}

\institute{University of Z\"urich, Winterthurerstrasse 190,
CH-8057 Z\"urich, Switzerland}

\maketitle

\begin{abstract}
We review the 3+1~-~split which serves to put Einstein's equations
into the form of a dynamical system with constraints. We then
discuss the constraint equations under the simplifying assumption
of time-symmetry. Multi-Black-Hole data are presented and more
explicitly described in the case of two holes. The effect of
different topologies is emphasized.
\end{abstract}

\subsubsection{Notation.} Space-time is a manifold $M$ with
Lorentzian metric $g$ of signature $(-,+,+,+)$. Greek indices 
are $\in\{0,1,2,3\}$ and latin indices are $\in\{1,2,3\}$.
Indices from the beginning of the alphabet, like $\alpha, 
\beta,\dots$ and $a,b,\dots$, refer to orthonormal frames and 
indices from the middle, like $\lambda,\mu,\dots$ and $l,m,\dots$ to 
coordinate frames. The symbol $\circ$ denotes the composition 
of maps. The relation $:=$ ($=:$) defines the left (right) 
hand side. The torsion and curvature tensors for the 
connection $\nabla$ are defined by $T(X,Y):=\nabla_XY-\nabla_YX-[X,Y]$ 
and $R(X,Y)Z:=\nabla_X\nabla_YZ-\nabla_Y\nabla_XZ-\nabla_{[X,Y]}Z$
respectively. The covariant components of the Riemann and Ricci 
tensors are defined by $R_{\alpha\beta\gamma\delta}:=
g(e_{\alpha},R(e_{\gamma},e_{\delta})e_{\beta})$ and 
$R_{\alpha\gamma}:=g^{\beta\delta}R_{\alpha\beta\gamma\delta}$
respectively. 

\section{The 3+1 -- Split}
In this article we discuss the vacuum Einstein equations
\begin{equation}
G^{\mu\nu}:=R^{\mu\nu}-\shalf g^{\mu\nu}R=0,
\label{1}
\end{equation}
which form a system of ten quasi-linear second order
differential equations for the ten functions $g_{\mu\nu}$.
However, the four equations $G_{\mu 0}=0$ do not involve the second
time derivatives and hence constrain the set of initial data.
To see this, recall that the twice contracted second Bianchi
identity gives $\nabla_{\mu}G^{\mu\nu}=0$, or expanded
\begin{equation}
\partial_0G^{0\nu}=-\partial_kG^{k\nu}
                   -\Gamma^{\mu}_{\mu\lambda}G^{\lambda\nu}
                   -\Gamma^{\nu}_{\mu\lambda}G^{\mu\lambda}.
\label{2}
\end{equation}
Since the right hand side contains at most second time derivatives
the assertion follows. The ten Einstein equations therefore split
into four {\it constraints} and six evolution equations $G^{ik}=0$.
That four equations constrain the initial data rather than
guiding the evolution results in four dynamically undetermined 
functions among the ten $g_{\mu\nu}$. The 
task is to parameterize the $g_{\mu\nu}$ in such a way that four 
dynamically undetermined functions can be cleanly separated
from the other six. How this can be done 
via the 3+1 split is explained below. The four dynamically 
undetermined quantities will be the famous  {\it lapse} 
(one function) and {\it shift} (three functions).
It follows directly from (\ref{2}) that the constraints will be
preserved under this evolution. 

The splitting of the Einstein equations will be formulated in a
geometric fashion. 
We initially think of $(M,g)$ as given and satisfying the Einstein
equations. Then we write down the evolution law for the intrinsic 
and  extrinsic geometry of a spacelike 3-manifold $\Sigma$ as it 
moves through $M$. Together with the constraints they are equivalent 
to all Einstein equations. Finally this procedure is turned upside 
down by taking the evolution equations for $\Sigma$'s geometry as 
starting point. Only after their integration can we construct 
the ambient space-time.

\subsection{3+1 Split Geometry}
The topology of space-time (or the portion thereof) which we want 
to decompose into space and time must be a product 
$M\cong\Sigma\times\reals$. We foliate $M$ by a one-parameter 
family of embeddings $e_t:\Sigma\rightarrow M$, $t\in\reals$.
For fixed $t$ the image of $e_t$ in $M$ is called $\Sigma_t$,
or the $t$'th leaf of the foliation. All leaves are assumed 
spacelike. Hence there is a normalized timelike vector field, $n$,
normal to all leaves. We choose one of the two possible orientations
and thereby introduce the notions of future and past: 
A timelike vector $X$ is future pointing iff $g(X,n)<0$
(recall signature convention). The tangent-bundle $T(M)$ can now be 
split into the orthogonal sum of the subbundle of spacelike 
vectors, $S(M)$, and the normal bundle, $N(M)$. The associated 
projection maps are given by 

\begin{eqnarray}
\S:\, T(M) & \rightarrow & S(M),\quad X\mapsto X+n\, g(n,X),\label{3}\\
\N:\, T(M) & \rightarrow & N(M),\quad X\mapsto  -n\, g(n,X),\label{4}
\end{eqnarray}
which can be naturally continued to the cotangent bundle by 
setting $\S^*(\omega):=\omega\circ\S$ and then factorwise 
on tensor products and linearly on the whole tensor-bundle. Thus 
we obtain a split of the whole tensor bundle, where from now on 
the projection maps are simply called $\S$ and $\N$ for all 
tensors. Tensors in the image of $\S$ are called {\it spatial}. 
It is easy to verify that
\begin{equation}
h:=\S g=g+\n\otimes\n ,
\label{5}
\end{equation}
where $\n:=g(n,\cdot)$. 
Note that the restriction $h_t$ of $h$ to $T(\Sigma_t)$ is just 
the induced Riemannian metric on $\Sigma_t$. Identifying for the 
moment $\Sigma$ and $\Sigma_t$ via $e_t$ this leads to $h_t=e_t^*g$ 
(Exercise). For what follows it is however crucial to regard 
spatial tensors as tensors over $M$ and not over $\Sigma$. 
Otherwise covariant (or Lie-) derivatives in directions off $\Sigma$ 
would not make sense.

If $X,Y$ are any spatial vector fields we can write
\begin{equation} 
\nabla_XY =  \S\nabla_XY + \N\nabla_XY = D_XY + n\,K(X,Y), 
\label{6}
\end{equation}
where we defined the spatial covariant derivative, $D$, and the 
extrinsic curvature, $K$, by 
\begin{eqnarray}
D_X    &:=&\S\circ\nabla_X, \label{7}\\
K(X,Y) &:=&-g(\nabla_XY,n)=-g(\nabla_YX,n)=g(\nabla_Xn,Y).\label{8}
\end{eqnarray}
The second equality in (\ref{8}) -- and hence the 
symmetry of $K$ -- follows from the vanishing torsion of $\nabla$ 
and the fact that $[X,Y]$ is spatial. It is easy to prove that
$K$ is indeed a tensor and that $D$ defines a connection 
on the tangent bundle of each leaf $\Sigma_t$. Extension 
via the Leibnitz rule leads to a unique connection on the bundle 
of spatial tensors, which can be directly defined by (\ref{7}) 
with the extended meaning of $\S$ described above. In fact it is 
just the Levi-Civita connection compatible with the metric $h$. 
To see this, we compute $D_Xh=\S\nabla_X(g+\n\otimes\n)=0$, since 
$\nabla_Xg=0=\S\n$, so that $D$ is compatible with $h$. 
Vanishing torsion is also immediate: 
$D_XY-D_YX-[X,Y]=\S (\nabla_XY-\nabla_YX-[X,Y])=0$, by 
$[X,Y]=\S [X,Y]$ and the vanishing torsion of $\nabla$. 

Let $\{e_0,e_1,e_2,e_3\}$ be an orthonormal frame adapted to the
foliation, i.e. $e_0=n$, and $\{e^0,e^1,e^2,e^3\}$ its dual. Then 
from (\ref{5}) with $\n=e^0$ we have
\begin{equation}
g=-e^0\otimes e^0+h=-e^0\otimes e^0+\sum_{a=1}^3e^a\otimes e^a.
\label{9}
\end{equation}
The family of embeddings $t\mapsto e_t$ defines a vector field, 
$\partial/\partial t=:\partial_t$, which is easily characterized 
by its action on any smooth function $f$:
\begin{equation} 
\partial_t f:=\frac{d}{dt}\Big\vert_{t=0}\, f\circ e_t.
\label{9a}
\end{equation}
This vector field can be decomposed into normal and tangential
components 
\begin{equation}
\partial_t=\alpha\, n+\beta=\alpha e_0+\beta^ae_a,
\label{9b}
\end{equation}
with uniquely defined function $\alpha$ and spatial vector field
$\beta$. They are called the {\it lapse} (function) and {\it shift}
(vector field) respectively. 

Let now $\{x^{\mu}\}$ be an adapted local coordinate system on $M$ 
so that $x^0=t$ and hence spatial fields 
$\partial_k:=\partial/\partial x^k$. The flow 
lines of $\partial_t$ are then the lines of constant spatial 
coordinates $x^k$. Hence $(\alpha, \beta)$ are interpreted as 
normal and tangential components of the 4-velocity -- measured in 
units of $t$ -- with which the points of constant spatial coordinates 
move. To express the metric $g$ in terms of these coordinates 
we use an obvious matrix notation and write 
\begin{eqnarray}
\pmatrix{\partial_t\cr\partial_k\cr}
&=&
\pmatrix{\alpha & \beta^a \cr
         0&A_k^a\cr}
\pmatrix{e_0\cr e_a\cr},
\label{10}
\\
\pmatrix{e^0&e^a\cr}
&=&
\pmatrix{dt & dx^k}
\pmatrix{\alpha&\beta^a\cr
         0&A_k^a\cr}.
\label{11}
\end{eqnarray}
Introducing (\ref{11}) into (\ref{9}) yields the 3+1 split
form of the metric $g$:
\begin{equation}
g=-\alpha^2\, dt\otimes dt+h_{ik}\,
   (dx^i+\beta^idt)\otimes (dx^k+\beta^kdt),
\label{12}
\end{equation}
where $h_{ik}=h(\partial_i,\partial_k)=\sum_a A^a_iA^a_k$ and 
$\beta^iA_i^a=\beta^a$. For the measure 4-form one easily
obtains $e^0\wedge e^1\wedge e^2\wedge e^3=
\alpha\sqrt{\det \{h_{ik}\}}d^4x$.

In the ambient space-time the notion of time-derivative of spatial 
tensors is introduced via the Lie-derivative along the time flow
generated by $\partial_t$. But in order to render this an
operation within the space of spatial tensor fields we must 
include a spatial projection. Using (\ref{9b}) we define
the ``doting'' by
\begin{equation}
\dot h:=\S L_{\partial_t}h=\alpha L_nh+\S L_{\beta}h,
\label{13}
\end{equation}
where we also used $L_{\alpha n}h=\alpha L_nh$ and that $L_nh$
is already spatial. This is true for any covariant spatial 
tensor and any smooth function $\alpha$. 
To prove this, we first remark that by Leibnitz' rule it 
suffices to prove it for a general spatial 1-form $\omega$. 
The first assertion now follows from $L_{\alpha n}\omega =
(i_{\alpha n}\circ d+d\circ i_{\alpha n})\omega=
\alpha i_n d\omega= \alpha L_n\omega$. The second statement follows 
from the general formula  $i_n\circ L_v=i_{[n,v]}+L_v\circ i_n$, 
showing that for $v=n$ the left hand side annihilates any spatial 
tensor field. This identity also shows why we need the projector in 
the second expression on the right hand side of (\ref{13}),
since for $v=\beta$ it shows that we would need 
$[n,\beta]\propto n$ for $L_{\beta}h$ to be spatial.
But this is generally false, as one easily shows that
$[n,\beta^k\partial_k]\propto n\Leftrightarrow 
\partial_t(\beta^k)=0$.

We proceed by showing that $L_nh$ is just twice the extrinsic 
curvature:
\begin{equation}
K=\shalf L_nh.
\label{14}
\end{equation}
To prove this relation, we take any spatial vector fields $X,Y$ and 
compute: $L_nh(X,Y)=\nabla_n(h(X,Y))-h([n,X],Y)-h(X,[n,Y])
=h(\nabla_Xn,Y)+h(X,\nabla_Yn)=2K(X,Y)$, where we used the 
metricity of $\nabla$, $(\nabla_nh)(X,Y)=(\nabla_ng)(X,Y)=0$,
and its vanishing torsion. Hence we arrive at
\begin{equation}
K=\frac{1}{2\alpha}\left(\dot h-\S L_{\beta}h\right).
\label{15}
\end{equation}
The projected Lie-derivative can be expressed in terms of the 
spatial covariant derivative in the usual way. In components with 
respect to a spatial coordinate frame this reads 
$(\S L_{\beta}h)_{ik}=D_i\beta_k+D_k\beta_i$.

\subsection{Constraints and Equations of Motion}
Using the splitting formula (\ref{6}) for the connection $\nabla$
in terms of $D$ and $K$ we can derive the so-called Gauss-Codazzi 
and Codazzi-Mainardi equations by a straightforward manipulation. 
In components with respect to $\{e_{\alpha}\}$ and with $\R3$ 
denoting the curvature of $D$, they read respectively: 
\begin{eqnarray}
R_{abcd}&=&\R3_{abcd}+K_{ac}K_{bd}-K_{ad}K_{bc},
\label{16}\\
R_{0abc}&=&D_cK_{ab}-D_bK_{ac}.
\label{17}
\end{eqnarray}
From here it is easy to write down the constraints by noting 
that in orthonormal frames one has 
$\sum_{a,b}R_{abab}=R+2R_{00}=2G_{00}$,  
i.e. the $00$ component of the Einstein tensor just depends on  
the spatial components of $\nabla$'s curvature. In fact, it is the 
sum of the spatial sectional curvatures of $\nabla$. Further,
$G_{0b}=R_{0b}=\sum_aR_{0aba}$. Hence we have the constraints,
now written in components with respect a coordinate frame,
\begin{eqnarray}
G_{\mu\nu}n^{\mu}n^{\nu}&=&\shalf(\R3-K_{ik}K^{ik}+(K_j^j)^2) = 0,
\label{18}\\
G_{\mu i}n^{\mu}&=&D^k(K_{ik}-h_{ik}K_j^j) = 0.
\label{19}
\end{eqnarray}

To obtain the dynamical equations one starts again from the 
defining equation of the curvature and manipulates the 
expression for $R_{0a0b}$.
Observing that $\nabla_n(g(e_a,\nabla_{e_b}n))=L_{n}K_{ab}$
one arrives at 
\begin{equation}
R_{0a0b}=-L_nK_{ab}+K_{ac}K^c_b + a_a a_b + D_aa_b,
\label{20}
\end{equation}
where $a:=\nabla_nn$. Note also that $a^{\flat}=L_nn^{\flat}$
(Exercise: Prove it).
Despite appearance, the last term in (\ref{20}) is also   
symmetric.\footnote{This is due to $n$ being 
hypersurface-orthogonal. To see this, we first note the identity  
$(L_n-\shalf i_n\circ d)dn^{\flat}\wedge n^{\flat}
= da^{\flat}\wedge n^{\flat}$ (Exercise: Prove it). 
Now, hypersurface-orthogonality of 
$n\Leftrightarrow dn^{\flat}\wedge n^{\flat}=0
\Rightarrow da^{\flat}\wedge n^{\flat}=0
\Leftrightarrow \S da^{\flat}=0\Leftrightarrow D_{[a}a_{b]}=0$.
}
Now, $R_{ab}=-R_{0a0b}+\sum_cR_{cacb}$, so that with (\ref{16})
we have
\begin{equation}
R_{ab}=\R3_{ab}+L_nK_{ab}+K_{ab}K^c_c-2K_{ac}K^c_b-a_aa_b-D_aa_b.
\label{21}
\end{equation}
This is almost the evolution equation we wish to obtain. 
As in (\ref{13}) we have $\dot K=\alpha L_nK+\S L_{\beta}K$, 
and since we want to write down the final equation in a coordinate 
basis, we can simplify the terms involving $a$ by noting that
$a_i=(L_nn^{\flat})(\partial_i)=
n^{\flat}([\partial_i,\frac{1}{\alpha}(\partial_t-\beta)])=
\partial_i\alpha/\alpha$. 
Hence\footnote{Be aware that some authors
define the extrinsic curvature with opposite sign, for example 
E.~Seidel in his lecture. Hence the discrepancy between our eqns. 
(17)(24) with his (5)(6) respectively. In our convention, which 
agrees with Hawking \& Ellis, a positive $K_i^i$ implies volume 
{\it expansion} under deformations in normal direction.
Also note that ``dotting" does not commute with index raising. 
Hence notations like ${\dot K}^{ij}$ are ambiguous. 
For example, denoting the index raising operation by a 
superscript $\sharp$, (\ref{14}) immediately gives
$(L_n(K^{\sharp})-(L_nK)^{\sharp})^{ik}=-4K^i_jK^{jk}$.}  
\begin{equation}
{\dot K}_{ik}=\alpha(2K_{ij}K^j_k-K_{ik}K^j_j+R_{ik}-\R3_{ik})
+L_{\beta}K_{ik}+D_iD_k\alpha ,
\label{22}
\end{equation}
where in the vacuum case we consider here one sets $R_{ik}=0$.
Note that in a coordinate frame dotting just means taking the 
partial derivative of the components, i.e., 
$L_{\partial_t}h_{ij}=\partial h_{ij}/\partial t$.

The dynamical formulation is now complete. 
The constraints are given by  eqs. (\ref{18})(\ref{19}) 
and the six evolution equations
of second order are written as twelve equations of first 
order, given by (\ref{13}) and (\ref{22}). The six dynamical 
components of $g$ are the $h_{ij}$, whereas there are no evolution 
equations for the four functions $\alpha,\beta$. The initial 
value problem thus takes the following form: 
1.)~choose a 3-manifold $\Sigma$ with local coordinates $\{x^i\}$,
2.)~find a Riemannian metric $h_{ij}$ and a symmetric covariant 
    tensor field $K_{ij}$ on $\Sigma$ which satisfy 
    (\ref{18})(\ref{19}),
3.)~choose any convenient functions $\alpha(t,x^k),\beta^i(t,x^k)$,
4.)~evolve $h_{ij}$ and $K_{ij}$ via (\ref{13})(\ref{22}) by using the
    choices made in the previous step,
5.)~take the solution curve $h_{ij}(t,x^k)$ and the functions from
    3.) to construct the space-time metric according to (\ref{12}),     
    where $x^0=t$. The $g$ so constructed solves Einstein's equations.
    An important theorem guarantees that for suitably specified data 
    a maximal evolution $(M,g)$ exists which is unique up to 
    diffeomorphisms (Choquet-Bruhat and Geroch 1969). See
    also Choquet-Bruhat and York~(1980) for a review and further
    references.

Regarding step 2.), we remark that all topologies $\Sigma$ allow
{\it some} initial data, i.e., there are no topological obstructions
to (\ref{18})(\ref{19}) (Witt 1986). This might change if 
geometrically {\it special} data are sought (see below).
To illustrate step 3.), we mention the so-called maximal slicing
condition on $\alpha$. To derive it, we compute $L_n(h^{ij}K_{ij})
=-2K^{ij}K_{ij}+h^{ij}L_nK_{ij}=-R_{00}-K^{ij}K_{ij}+\Delta\alpha/\alpha$,
where $\Delta=D^iD_i$ and where we used $R_{00}=\sum_aR_{0a0a}$ and 
(\ref{20}) to replace $L_nK$. Hence
\begin{equation}
L_{\partial_t}(K^i_i)=
(\Delta-K^{ij}K_{ij}-R_{00})\alpha +L_{\beta}(K^i_i),
\label{22a}
\end{equation}
where we left in $R_{00}$ for generality. Note that the strong energy 
condition implies $R_{00}\geq 0$ through the Einstein equations. 
The hypersurface $\Sigma\subset M$ is called {\it maximal} if the 
trace of its extrinsic curvature --~the so-called 
{\it mean curvature}~-- is zero, i.e., $K_i^i=0$. This is equivalent to 
$\Sigma$ being a stationary point of all the 3-dimensional volume
functionals for domains in $\Sigma$ and fixed boundaries.
\footnote{The standard terminology is that such stationary points 
are called ``maximal'' if the ambient geometry is Lorentzian
and ``minimal'' if it is Riemannian, irrespectively of whether 
they really are true maxima or minima respectively. True extrema 
are called stable maximal (minimal) surfaces.}
(Exercise: Prove this using (\ref{14}).) 
Now, given a maximal slice $\Sigma\subset M$, (\ref{22a}) gives 
the following simple condition on $\alpha$ if the evolution is
to preserve maximality: ${\cal O}\alpha=0$ with elliptic
operator ${\cal O}=\Delta-K^{ij}K_{ij}-R_{00}$. 
(Exercise: Assuming the strong energy condition, prove that any 
smooth function $\alpha$ in the kernel of ${\cal O}$ cannot have a 
positive local maximum or negative local minimum on $\Sigma$.)
In the vacuum case one can use (\ref{18})
to write ${\cal O}=\Delta-\R3$, i.e., purely in terms of the 
intrinsic geometry of $\Sigma$, where clearly $\R3\geq 0$.

The maximal slicing condition plays an important r\^ole in numerical
evolution schemes, since --~by definition~-- the evolving maximal 
slices $\Sigma_t\subset M$  approach slowest the regions of strongest 
spatial compression. In this sense they have the tendency to avoid 
singularities. For further information see section 2.3 of 
E.~Seidel's lecture. Finally we note that since not all topologies 
$\Sigma$ allow for metrics with $\R3\geq 0$, there exist topological 
obstructions to {\it maximal} initial data sets (Witt 1986).

\section{Time-Symmetric Initial Data}
Suppose a hypersurface $\Sigma\subset M$ has vanishing extrinsic
curvature, $K=0$. From (\ref{6}) we then have $\nabla_XY=D_XY$ for 
all vector fields $X,Y$ tangent to $\Sigma$. In particular, if
$\gamma:I\rightarrow\Sigma$ is a curve with tangent vector field 
$\gamma'$ over $\gamma$, then 
$\nabla_{\gamma'}\gamma'=D_{\gamma'}\gamma'$ and $\gamma$
is a geodesic in $\Sigma$ iff it is a geodesic in $M$.
Submanifolds for which this is true are called {\it totally geodesic}.
This is a stronger condition than maximality.
In general, constant mean curvature data play an important
r\^ ole in the solution theory for the constraints
(see York 1973, \'O Murchadha and York 1974).
Here we shall vastly shortcut the general procedure by imposing the 
condition that $\Sigma$ be totally geodesic. One can then show that 
the maximal development, $M$, from these data allows an isometry
fixing $\Sigma$ pointwise and exchanging the two components of 
$M-\Sigma$. Hence such data are called {\it time symmetric}.
For such cases the constraints reduce to the simple condition that 
$(\Sigma,h)$ has vanishing Ricci-scalar:
\begin{equation}
\R3(h)=0,
\label{23}
\end{equation}
where for later convenience we explicitly indicated the metric 
as argument of $\R3$. A general idea for solving (\ref{23}) is to  
prescribe $h$ up to an overall conformal factor $\Phi$, 
and let (\ref{23}) determine the latter. So setting
$h=\Phi^4 h'$, with fourth power just for convenience, we have by
the conformal transformation law for the Ricci-scalar
\begin{equation}
\R3(\Phi^4 h')
=-8\Phi^{-5}(\Delta_{h'}-\eights \R3(h'))\Phi
=:-8\Phi^{-5}\,C_{h'}\Phi=0,
\label{24}
\end{equation}
where $\Delta_{h'}$ is the Laplacian for the metric $h'$.
We are interested in $C^2$ solutions satisfying $\Phi>0$
and where $(\Sigma,h)$ has no boundaries at finite distances,
i.e. $\Sigma$ should be topologically complete in the metric
topology defined by the distance function induced by $h$.
The last condition is equivalent to $(\Sigma,h)$ being geodesically
complete (theorem of Hopf-Rinow-DeRahm, see e.g. Spivak 1979).
In addition, we shall only be interested in manifolds
whose ends are asymptotically flat.
Allowing the manifold $\Sigma$ to have more ends or to be
otherwise topologically more complicated allows for a greater variety
of solutions. Note that to each of $n$ asymptotically flat
ends there corresponds an ADM-mass of which $n-1$ are 
independent (see below).

\subsubsection{Brill Waves.} One may ask whether simple
asymptotically flat solutions to $C_{h'}\Phi=0$ exist on
$\Sigma=\reals^3$. There are no (regular!) black-hole
solutions with this simple topology, but there
are solutions representing localized gravitational 
waves of non-zero total ADM energy (Araki 1959).
In the axisymmetric case they were investigated in detail by 
Brill (1959). 
Solutions of this kind are collectively called ``Brill waves''.
One takes (from now on in the usual shorthand suppressing the 
$\otimes$)
\begin{equation} 
h'=\exp(\lambda\, q(z,\rho))(dz^2+d\rho^2)+\rho^2\,d\varphi^2 ,
\label{25}
\end{equation}
where the profile-function $q$ must for $r\rightarrow\infty$
fall off like $r^{-2}$ and like $r^{-3}$ in its first 
derivatives in order for $h$ to turn out asymptotically flat. 
$q$ characterizes the geometry in the meridial
cross section ($z\rho$-plane) of the toroidal gravitational wave. 
Regularity on the axis also requires $q$ and 
$\partial_{\rho}q$ to vanish for $\rho=0$. 
The parameter $\lambda\in\reals_+$ is sometimes introduced to 
independently parameterize the overall amplitude. Equation 
(\ref{23}) for $\Phi(z,\rho)$ takes the particularly simple form 
\begin{equation}
\left(\Delta_{\f} + \quarter\lambda \Deltatwo q\right)\Phi=0,
\label{26}
\end{equation}
where $\Delta_{\f}$ is the {\it flat} Laplacian and 
$\Delta^{(2)}=\partial^2/\partial z^2+\partial^2/\partial\rho^2$.
Given $q$, everywhere positive solutions for $\Phi$ exist 
provided $\lambda$ is below some critical value depending on 
the choice $q$ (Araki 1959). To see uniqueness, assume the existence
of two solutions $\Phi_1$ and $\Phi_2$ and set $h_i=\Phi^4_ih'$,
$i=1,2$. Then $\Phi_3:=\Phi_1/\Phi_2$ is also $C^2$, positive and 
tends to 1 at infinity. But (\ref{24}) immediately implies 
$C_{h_2}\Phi_3=-\eights\Phi_3^5\R3(h_1)=0$, and since also 
$\R3(h_2)=0$ this is equivalent to $\Delta_{h_2}\Phi_3=0$. Hence
$\Phi_3=1$ due to the fact that the only bounded harmonic 
functions are the constant ones.

\section{Black-Hole Data}
A substantial variety of time-symmetric black-hole data can already
be obtained by solving (\ref{24}) when $h'$ is {\it flat}, i.e.,
where the 3-metric, $h$, on the spatial slice at the moment of
time-symmetry is conformally flat. One can obtain manifolds
with any number of asymptotically flat ends, and then reduce this
number by a process which is best described by calling it ``plumbing"
(see below). We shall devote the rest of this paper to the description 
of such solutions and techniques. Note that for flat $h'$ we are left 
with the simple harmonic equation involving only the flat Laplacian:
\begin{equation}
\Delta_{\f}\Phi=0.
\label{27}
\end{equation}

In general it is difficult to infer from given initial data whether 
they correspond to a spacetime with black holes, i.e. with event 
horizons. However, in the examples to follow it is easy to see that 
that there will be apparent horizons, since for time symmetric data
apparent horizons correspond precisely to minimal surfaces 
$S\subset\Sigma$ \footnote{The condition on $S$ being an apparent 
horizon is that the congruences of outgoing null rays from $S$
must have zero divergence. Analytically this translates into 
$\hbox{tr}_2(\kappa)=\pm(\hbox{tr}(K)-K(\nu,\nu)$ where
$\kappa,\nu$ are respectively the extrinsic curvature and normal 
of $S$ in $\Sigma$. The upper sign is valid for {\it past} 
apparent horizons, and the lower one for {\it future} apparent 
horizons. $\hbox{tr}_2$ is the 2-dimensional trace using the    
induced metric of $S$ and $\hbox{tr}$ the 3-dimensional trace 
using $h$. For time-symmetric initial data ($K=0$) 
this condition states that $\kappa$ is traceless and hence $S$ 
minimal in $\Sigma$.}. 
Proposition 9.2.8 of Hawking and Ellis~(1973) now implies the 
existence of an event horizon whose intersection with $\Sigma$ is 
on, or outside, the outermost apparent horizon for any regular 
predictable spacetime that develops from data satisfying the strong 
energy condition. Concerning the topology of apparent horizons
we remark the following: Using the formula for the second variation 
of the area functional and the theorem of Gau\ss\ -Bonnet, one shows  
that for ambient metrics with non-negative Ricci scalar any 
connected component of an orientable stable minimal surface 
of finite volume must be a topological 2-sphere (Gibbons 1972). 
Allowing also for non-orientable apparent horizons, one deduces 
from this that for metrics $h$ of $\Sigma$ with $\R3(h)\geq 0$ a 
connected component of an apparent horizon is either $S^2$ or $RP^2$, 
the latter being the (non-orientable) 2-dimensional real projective 
space. If $\Sigma$ is orientable $RP^2\subset\Sigma$ is one-sided, 
as in the example below.

\subsection{Schwarzschild Data}
We start by noting that the most general non-trivial 
solution of (\ref{27}) on $\Sigma=\reals-\{0\}$ is given by
$\Phi(\vec x)=1+\frac{m}{2r}$ with $r=\Vert\vec x\Vert$ and
$m\in\reals_+$.
We cannot have any higher multipole moments because
then $\Phi$ necessarily has zeros on $\Sigma$. 
Just removing $\Phi^{-1}(0)$ from $\Sigma$ does not work 
since these points are at finite distance so that the resulting
space would not be (geodesically) complete. This is also
the reason why $m$ must be positive. Hence we obtain for 
the metric $h$ in polar coordinates 
\begin{equation}
h=\left(1+\frac{m}{2r}\right)^4\,(dr^2+r^2\,d\Omega^2),
\label{28}
\end{equation}
with $d\Omega^2=d\theta^2+\sin^2\theta\,d\varphi^2$.
Now, it is easy to verify that the following two diffeomorphisms, 
$I$ and $\t I$, of $\Sigma$ are involutive (i.e., square to the identity) 
isometries:
\begin{eqnarray}
I(r,\theta,\varphi)&:=&
\left(\frac{m^2}{4r},\theta,\varphi\right),
\label{29}\\
\t I(r,\theta,\varphi)&:=&
\left(\frac{m^2}{4r},\pi-\theta,\varphi+\pi\right).
\label{30}
\end{eqnarray}
The map $I$ is called an inversion on the sphere $r=m/2$, whereas 
$\t I$ is that inversion plus an additional antipodal map on the 
spheres of constant $r$. We shall sometimes refer to them as 
inversions of the first and second kind respectively. 
$\t I$ has no fixed points while $I$ fixes each point of the sphere 
$S=\{\vec x \mid r=\frac{m}{2}\}$ (which, as set, is also left 
invariant by $\t I$). 
As fixed point set of an isometry $S$ must be totally 
geodesic\footnote{Proof: Consider the unique geodesic $\gamma$ 
starting on and tangentially to $S$. It cannot leave $S$ since if 
it would, its image under $I$ would be a {\it different} geodesic 
with the same initial conditions, which contradicts the uniqueness 
theorem for ODE's.}, hence minimal and therefore an apparent 
horizon. Its surface area is $A=16\pi m^2$, and it separates the 
two isometric regions $r>m/2$ and $r<m/2$.
%
%
The metric (\ref{28}) corresponds to the spatial part
of the Schwarzschild metric of mass $m$ in isotropic coordinates,
which cover both asymptotically flat regions (I and III) on the 
Kruskal manifold. Using this isotropic form, one can read off 
$\alpha=(1-m/2r)/(1+m/2r)$, $\beta=0$ and verify that with this 
choice the static form of (\ref{22}) with $K=0$ is satisfied 
(Exercise). 

The manifold $\Sigma$ has two isometric ends and we can get 
rid of one by suitably identifications. For this we take the   
quotient $\tilde\Sigma$ of $\Sigma$ with respect to the free
action of $\t I$. The freeness guarantees that the quotient will be  
a manifold, and, by being an isometry, the metric descends to  
a smooth metric on the quotient. $\tilde\Sigma$ can be pictured 
by cutting $\Sigma$ along $S$, throwing away one piece, and 
identifying opposite points on the inner boundary $S$ on the 
retained piece. Hence topologically $\tilde\Sigma$ is the 
real projective space, $\reals P^3$, minus a point. The projection of 
$S$ into $\tilde\Sigma$ is a totally geodesic, one-sided 
(i.e. non-orientable) surface diffeomorphic to $\reals P^2$. 
$\tilde\Sigma$ is orientable, smooth,
complete and with one end which is isometric to, and hence has the 
same ADM mass as, either end in $\Sigma$. This demonstrates how
the introduction of more ends or other topological features 
makes it possible to define non-trivial black-hole data.
One may also combine Brill waves with a black hole to model
a single distorted black hole. This is further discussed in 
section 2.2 of E.~Seidel's lecture.

\subsubsection{Multi-Schwarzschild Data.}
Taking $\Sigma=\reals^3-\{\vec c_1,\cdots,\vec c_n\}$ the 
generalization of (\ref{28}) is easily obtained
with $n$ poles of strengths $a_i\in\reals_+$ at 
``positions'' $\vec c_i$:
\begin{equation}
\Phi(\vec x)=1+\sum_{i=1}^n\frac{a_i}{r_i},
\label{31}
\end{equation}
where $r_i:=\Vert\vec x-\vec c_i\Vert$.
For each $i$ we can introduce inverted polar coordinates 
$r'_i=a_i^2/r_i$ to probe the region $r_i\rightarrow 0$
by letting $r'_i\rightarrow\infty$. Doing this shows that the 
metric is asymptotically of the form (\ref{28}) with certain 
mass parameters $m=m_i$ given below. The same is true for 
the region $r\rightarrow\infty$ with mass $M$. Hence one 
obtains $n+1$ asymptotically flat ends. 
%
%
The internal masses and the overall mass are given by 
($r_{ji}:=\Vert\vec c_j-\vec c_i\Vert$)
\begin{equation}
m_i=2a_i(1+\chi_i),\quad\hbox{where}\quad
\chi_i:=\sum_{j\not =i}\frac{a_j}{r_{ji}},
\quad\hbox{and}\quad
M=2\sum_ia_i. 
\label{32}
\end{equation}
In terms of the parameters $a_i,r_{ij}$ the binding energy takes 
the simple form
\begin{equation}
\Delta M:=M-\sum_{i=1}^n m_i=-2\sum_{i=1}^n a_i\chi_i
=-2\sum_{i=1}^n\sum_{j\not =i}\frac{a_ia_j}{r_{ij}} < 0.
\label{33}
\end{equation}
Note that there are as many independent masses as there are
generators of the second homology group of $\Sigma$. These
generators may be represented by stable minimal surfaces associated to
each internal end. Their surfaces areas clearly satisfy 
$A_i>16\pi (2a_i)^2$, since the right hand side represents 
the minimal area in the strictly smaller metric (31) for 
just one hole with parameter $m=2a_i$. But there is also an upper 
bound for the area, given by the recently proven Riemannian Penrose 
inequality\footnote{The proof of \cite{Hui} applies to all 
asymptotically flat Riemannian 3-Manifolds whose Ricci scalar 
satisfies $R\geq 0$. They prove that the area $A$ of the outermost 
stable minimal surface bounding an end and the ADM mass $m$ of that 
end satisfy $A\leq 16\pi m^2$. It implies the positive mass theorem
for data with $R\geq 0$.} (Huisken and Ilmanen 1997), which in our 
context reads $A\leq 16\pi (2a_i)^2(1+\chi_i)^2$. Assuming 
the existence of an event horizon (see above), the Area Theorem 
(see my other contribution to this volume) implies that the 
area of the hole in the $i$'th end cannot evolve below $A_i$, 
which, using the first inequality above, implies in particular 
that the energy which is bound in the final hole is greater than 
$2a_i=m_i/(1+\chi_i)$. The difference of the (conserved) ADM mass 
$m_i$ to the mass of the final hole is therefore bounded above 
by $m_i\chi_i/(1+\chi_i)$. In other words, the fraction of energy 
being radiated is less than $\chi_i/(1+\chi_i)$.
This still allows for total conversion into radiation 
if one chooses $\chi_i\rightarrow\infty$. 

It would be of course more interesting to express (\ref{33}) in terms
of physical variables, like the individual masses $m_i$, and more 
geometrically defined distance functions than $r_{ij}$, like 
e.g. the proper geodesic distance of the minimal surfaces in 
the $i$'th and $j$'th throat.
Note that for small mass-to-separation ratios we may in a first 
approximation replace $a_i$ by $\shalf m_i$ and $r_{ij}$ by the 
geodesic distance of the $i$'th and $j$'th apparent horizons 
and get the familiar Newtonian formula. But there will be 
corrections the precise form of which depend on ones 
definition of ``distance between two holes''. 
Whereas here mass is unambiguously defined for each hole 
(by ADM), there is no natural definition of distance.
Perhaps the easiest intrinsically defined distance is the one
given above. For the multi-Schwarzschild manifold
it has the disadvantage that the minimal surfaces are not 
easy to locate analytically and one has to resort to numerical methods 
(see Brill and Lindquist 1963 for early attempts). 

The location of minimal surfaces is interesting for a variety of 
reasons. It somewhat simplifies in the case of just two holes, 
which is automatically axisymmetric. Then the variational 
principle for the minimal surfaces reduces to a geodesic 
principle for curves in the $z\rho$-half-plane (cylindrical 
coordinates). The appropriately parameterized solution curves 
just describe a motion of a point particle in the potential 
$-\shalf\rho^2\Phi^8$ (\v Cade\v z 1974).
However, general analytic solution still do not exist. Numerical
studies by \cite{Bishop} for equal masses ($a_1=a_2=:a$) show the
very interesting behaviour above the critical value $a/r_{12}\simeq
1/1.53$, where {\it two} more minimal surfaces appear, each of which
enclosing the previous two. Initially they coincide, but for
increasing $a/r_{12}$ they separate with the inner one rapidly
increasing in area whereas the outermost staying almost constant.
See also \cite{Gibbons2} for a related discussion.
%
%

For the data discussed below the difficulty of determining location and
size of minimal surfaces is absent, but somewhat as trade-off the concept
of individual mass now becomes slightly more problematic.

\subsubsection{Different Topologies for Multi-Hole Data.}
There are other generalizations of the single hole case. 
The ones we discuss now will preserve the existence of 
involutive isometries like (\ref{29}-\ref{30}), but now for 
{\it each} apparent horizon. The manifolds they exist on 
have two or even just one end. The construction is somewhat 
involved (Lindquist 1963) and uses the method of images
to construct solutions to (\ref{27}). This method was introduced
by \cite{Misner2} for the time symmetric case and later generalized
to more general situations (e.g. Bowen and York 1980, Bowen 1984).
(There is also  a recent alternative proposal by \cite{BB}.)
For the general understanding it will be sufficient to 
explain the construction for just two holes.
Note that the ADM definition of mass cannot be applied to 
the individual hole if it does not have an asymptotically 
flat end associated to it. But there exist alternative
proposals for mass due to Lindquist~(1963) and Penrose~(1982)
which can be employed here. (See also the general review 
by Penrose (1984).) But it should be pointed out
that these definitions do not always apply in more general
situations. For example, for the applicability of Penrose's
mass definition within time-symmetric hypersurfaces the metric of
this hypersurface must be conformally flat (Tod~1983, Beig~1991).

\subsection{Two Hole Data}
Just as in electrostatics, we shall use the method of images to 
construct special solutions to (\ref{27}). This is done by placing 
image masses in an auxiliary, fictitious space so as to enforce 
special properties of $\Phi$. The properties which will be enforced 
here are such that the inversions (\ref{29})(\ref{30}) on 
2-spheres become isometries.

We start by drawing two 2-spheres 
$S_i:= S(a_i,{\vec c}_i)$, $i=1,2$, with radii $a_i$ and centered 
at ${\vec c}_i$. The spheres are non-intersecting and outside each 
other, so that $r_{12}>a_1+a_2$. On $\reals^3-\{{\vec c}_i\}$ we 
have the diffeomorphisms $I_i$ and ${\t I}_i$, which in polar 
coordinates at ${\vec c}_i$ take the forms (\ref{29}) and (\ref{30}) 
respectively. These induce involutions on the space of 
functions, defined by  
\begin{equation}
J_i(f):=\frac{a_i}{r_i}\, f\circ I_i\quad\hbox{and}\quad
{\t J}_i(f):=\frac{a_i}{r_i}\, f\circ {\t I}_i
\label{34}
\end{equation}
respectively, where $f$ is any function. The crucial property of 
these maps is 
\begin{equation}
\Delta_{\f}\circ J_i=(a_i/r_i)^4\, J_i\circ\Delta_{\f}\quad\hbox{and}\quad
\Delta_{\f}\circ{\t J}_i=(a_i/r_i)^4\,{\t J}_i\circ\Delta_{\f},
\label{35}
\end{equation}
which in particular implies that the image of a harmonic function 
will again be harmonic, although with different singularity structure. 
The image of the constant function, 
$f\equiv 1$, under either of these maps is just 
$f'=a_i/r_i$, i.e., the pole of strength $a_i$ at $\vec c_i$. 
Moreover, given the unit pole $f(\vec x)=1/\Vert\vec x-\vec d\Vert$
at $\vec d$ outside $S_i$, then its image under $J_i$ is 
\begin{equation}
J_i(f)=\frac{a_i}{\Vert{\vec c}_i-\vec d\Vert}\,
       \frac{1}{\Vert\vec x-I_i(\vec d)\Vert},
\label{36}
\end{equation}
and correspondingly for ${\t J}_i$. It represents a pole of
strength $a_i/\Vert{\vec c}_i-\vec d\Vert<1$ at the image point
$I_i(\vec d)$ (resp. ${\t I}_i(\vec d)$). 

Writing down the metric $h=\Phi^4\,ds^2_{\f}$ in polar 
coordinates centered at $\vec c_i$ one easily verifies that 
$I_i$ (${\t I}_i$) is an isometry of $h$ if $\Phi$ is 
invariant under $J_i$ (${\t J}_i$). The construction of such an  
invariant $\Phi$ is by brute force: One averages the function 
$\Phi_0\equiv 1$ over the free product of the groups generated by 
$J_1,J_2$ (${\t J}_1,{\t J}_2$). The elements of this 
free-product-group are strings of alternating $J_1$'s and $J_2$'s,
where for each string length $n\geq 1$ there are the two different 
elements $J_1\circ J_2\circ J_1\cdots$ and $J_2\circ J_1\circ J_2\cdots$.
By definition, the string of length $0$ is the identity element. 
Hence one sets
\begin{equation}
\Phi_N:=1+\sum_{n=1}^N\sum J_{i_1}\circ \cdots\circ J_{i_n}\,(\Phi_0),
\label{37}
\end{equation}
where the first sum is over the two different elements of 
length $n$.
On $\reals^3-\{\hbox{image points}\}$ the sequence $\Phi_N$ 
converges to a smooth function $\Phi$ for $N\rightarrow\infty$.
Convergence follows because at level $N$ the strengths of the new 
poles are suppressed by at least a factor of $q^{N-1}$, where   
$q=\hbox{sup}_{i,j}\,a_i/(r_{ij}-a_j)<1$. Note also that all image 
poles in $S_i$ lie in fact in the interior of the concentric 
but smaller sphere of radius $a'_i:=a_i^2/(r_{ij}-a_j)$.
Cutting out the interiors of $S(a'_i,\vec c_i)$ $i=1,2$ 
thus leaves the spheres $S_i$ with small collar neighborhoods 
the two sides of which are isometrically mapped into each other
by $I_i$ (or ${\t I}_i$). Using two copies of the manifold so 
obtained we can pairwise identify these collar neighborhoods using
these isometries so that an Einstein-Rosen manifold with two
bridges results. 
Their topology is that of the twice punctured 
``handle'' $S^1\times S^2$ with each puncture corresponding to  
an asymptotically flat end. This construction generalizes to any 
number $N$ of holes (or bridges), where as manifold one obtains the
twice punctured connected sum of $N-1$ handles.
(For the notion of connected sums see e.g. Giulini 1994.)
%
%
For two holes of equal mass one may also just identify $S_1$ and
$S_2$ and get Misner's wormhole (Misner 1960) if one uses inversions 
of the first kind, or its non-orientable counterpart if one uses 
inversions of the second kind (Giulini 1990). 
%
%
Both manifolds just have one end. In the second case one has the 
additional possibility to just close ``close-off'' the
spheres $S_i$ individually by identifying its antipodal points using
${\t I}_i$ (Giulini 1992). The manifold has the topology of the once
punctured connected sum of two real projective spaces $\reals P^3$.
It is orientable and has only one asymptotically flat end. It can
be seen as the generalization to two holes of the once punctured
single $\reals P^3$ obtained above. This construction also generalizes
to any number $N$ of holes and one obtains the once punctured 
connected sum of $N$ $\reals P^3$'s. These manifolds are doubly  
covered by the $N$-bridge manifolds discussed above.

\subsection{Analytic Expressions}
In the case of two holes there exists a geometrically adapted 
coordinate system -- so called spherical bi-polar coordinates --
which allows to write down explicit expressions. We take 
$a_1=a_2=a$, ${\vec c}_1=d\,{\vec e}_z$ and 
${\vec c}_2=-d\,{\vec e}_z$.
Taking the $a_i$'s equal means that the holes are of equal size 
(individual mass). We thus consider a two parameter family of  
configurations labeled e.g. by mass (overall or individual)
and separation. All image poles are on the $z$-axis whose 
strengths $a_n$ and locations $d_n$ (positively counted $z$ 
coordinate) satisfy the coupled recursion relations 
\begin{equation}
a_n=a_{n-1}\frac{a}{d+d_{n-1}},\qquad
d_n=d\mp\frac{a^2}{d+d_{n-1}},
\label{38}
\end{equation}
where the upper (lower) sign is valid for inversions of the first 
(second) kind. Using instead of $a,d$ the parameters  
$c,\mu_0$ defined by $a:=c/\sinh \mu_0$, $d:=c\,\coth \mu_0$
we can solve the recursion relations by
\begin{equation}
a_n=\frac{c}{\sinh n\mu_0},\qquad
d_n=c\,\coth n\mu_0,
\label{39}
\end{equation}
for the upper sign, and for the lower sign
\begin{eqnarray}
a_n&=&\frac{c}{\sinh n\mu_0},\qquad
d_n = c\,\coth n\mu_0\qquad\hbox{for $n$ even,}
\label{40}\\
a_n&=&\frac{c}{\cosh n\mu_0},\qquad
d_n = c\,\tanh n\mu_0\qquad\hbox{for $n$ odd.}
\label{41}
\end{eqnarray}
In the $xz$-plane we introduce bi-polar coordinates via 
$\exp(\mu-i\eta)=(\xi+c)/(\xi-c)$ with $\xi=z+ix$. By construction 
the lines of constant $\mu$ intersect those of constant $\eta$ 
orthogonally. Both families consist of circles; those in the 
first family are centered on the $z$-axis with radius $c/\sinh \mu$ 
at $\vert z\vert=c\,\coth \mu$, and those in the second family on 
the $x$-axis with radius $c/\sin \eta$ at $\vert x\vert=c\,\cot \eta$. 
%
%
Rotating this system around the $z$-axis with azimuthal angle $\phi$ 
leads to the spherical bi-polar coordinates. Explicitly one obtains 
\begin{equation}
x=c\frac{\sin\eta\cos\phi}{\cosh\mu-\cos\eta},\quad
y=c\frac{\sin\eta\sin\phi}{\cosh\mu-\cos\eta},\quad
z=c\frac{\sinh\mu}{\cosh\mu-\cos\eta}.
\label{42}
\end{equation}
Together with (\ref{39}-\ref{41}) this gives
\begin{equation}
\frac{a_n}{\Vert\vec x\pm d_n{\vec e}_z\Vert}
=\frac{[\cosh\mu-\cos\eta]^{1/2}}
      {[\cosh(\mu\pm 2n\mu_0)-\varepsilon\,\cos\eta]^{1/2}},
\label{43}
\end{equation}
where $\varepsilon=1$ if one uses inversions of the first kind and 
$\varepsilon=-1$ if one uses those of the second kind.
The final expression for the metric in $(\mu,\eta,\phi)$-coordinates 
can now be written down:
\begin{eqnarray}
h&=&\left[1+\sum_{n=1}^{\infty}
\left(\frac{a_n}{\Vert\vec x+d_n{\vec e}_z\Vert}
     +\frac{a_n}{\Vert\vec x-d_n{\vec e}_z\Vert}
\right)\right]^4\,d\vec x\cdot d\vec x
\\
&=&\left[\sum_{n\in\integers}\left(\cosh(\mu+2n\mu_0)
-\varepsilon^n\,\cos\eta\right)^{-1/2}\right]^4\,
(d\mu^2+d\eta^2+\sin^2\eta\,d\phi^2).\qquad
\label{44}
\end{eqnarray}
It nicely exhibits the isometries
$(\mu,\eta,\phi)\mapsto(\mu+2\mu_0,\eta,\phi)$ for $\varepsilon =1$ and
$(\mu,\eta,\phi)\mapsto(\mu+2\mu_0,\pi-\eta,\phi)$ for $\varepsilon=-1$.
The extrinsic curvature matrix for the surfaces of constant $\mu$ with 
respect to an orthonormal basis in $\eta$ and $\phi$ direction is 
given by $2\Phi^{-3}\partial\Phi/\partial\mu$ times the unit
matrix. Hence $K$ has only a trace part (the surfaces of constant
$\mu$ are totally umbillic) and vanishes iff $\mu=\pm\mu_0$.
Hence in both cases, $\varepsilon=\pm 1$, the apparent horizons are
also totally geodesic (this we already knew for $\varepsilon=1$).

Next we turn to the expressions for the masses. We shall follow
Lindquist (1963) and define the mass of the first hole 
by appropriately applying (\ref{32}): We sum all the ``bare masses'' 
$2a_i$ in $S_1$, each enhanced by an interaction factor 
$1+\chi'_i$ which includes the interactions of each pole
in $S_1$ with any pole in $S_2$, but {\it not} with poles 
in $S_1$. This we write as

\begin{equation}
m_1=2\sum_{i\in S_1}a_i
\left(1+\sum_{j\in S_2}\frac{a_j}{r_{ij}}\right),
\label{45}
\end{equation}
with the obvious meaning of ``$\in$''.
Since $m_1=m_2$ we write $m$ for the individual mass and $M$ for the  
overall mass. The latter is just the sum of all $2a_i$. Using 
(\ref{39}-\ref{41}) one obtains (quantities referring to 
$\varepsilon=-1$ carry a tilde)
\begin{equation}
m=2c\sum_{n=1}^{\infty}\frac{n}{\sinh n\mu_0},\qquad
M=4c\sum_{n=1}^{\infty}\frac{1}{\sinh n\mu_0},
\label{46}
\end{equation}
for $\varepsilon=1$, and for $\varepsilon=-1$
\begin{eqnarray}
\t m&=&  2c\sum_{n=1}^{\infty}\frac{2n}{\sinh 2n\mu_0}
      +2c\sum_{n=0}^{\infty}\frac{2n+1}{\cosh (2n+1)n\mu_0},\qquad       
\label{47}\\
\t M&=&  4c\sum_{n=1}^{\infty}\frac{1}{\sinh 2n\mu_0}      
      +4c\sum_{n=0}^{\infty}\frac{1}{\cosh(2n+1)n\mu_0}.
\label{48}
\end{eqnarray}

As mentioned above, we define the distance of the holes as the  
geodesic distance of the apparent horizons $\mu=\pm\mu_0$. 
The shortest geodesic connecting these two surfaces is $\eta=\pi$.
For $\varepsilon=1$  its length, $\l$, may be expressed in closed 
form:
\begin{equation}
l=2c(1+2m\mu_0),
\label{49}
\end{equation}
with $m$ from (\ref{46}). I have not been able to find such a compact 
expression in the case $\varepsilon=-1$. 

Like $\t l$, many quantities of interest cannot be evaluated  
in closed form. In these cases it may be useful to expand 
in powers of $m/l$. Numerical studies show that additional outer 
apparent horizons form (i.e. the holes merge) for values above 
$m/l\simeq 0.26$ (Smarr et al 1976), so that good convergence holds
up to the merging ratio.

\subsubsection{Comparing $\varepsilon=1$ to $\varepsilon=-1$.}
We have seen that mathematically these two cases differ by allowing 
different topologies. But are there more physical aspects in 
which they differ? A natural question is how for fixed 
``physical'' variables $m=\t m$ and $l=\t l$ the total energies 
$M$ and $\t M$ differ (Giulini 1990). One finds
\begin{equation}
\frac{\t M-M}{M}=-\left(\frac{m}{2l}\right)^2 +{\cal O}(3)\,,
\label{50}
\end{equation}
showing that for $\varepsilon=-1$ the holes are slightly tighter 
bound (i.e. they attract stronger), although the additional energy gained  
until merge is only about $10^{-2}M$. This result is qualitatively 
unchanged if one uses Penrose's instead of Lindquist's definition  
of mass.

Another difference shows up in the deformation of the apparent 
horizons upon (adiabatic) approach of the two holes. One can 
define an intrinsic deformation parameter as follows:
Regard $(\eta,\phi)$ as polar coordinates. The poles are the zeros 
of the Killing field $\partial_{\phi}$. Define  
$C_{\eta}$ as twice their geodesic distance. Among the orbits of 
$\partial_{\phi}$ is one of greatest length, $C_{\phi}$. The 
deformation parameter is $D:=(C_{\eta}-C_{\phi})/C_{\eta}$.
One obtains (Giulini 1990)
\begin{eqnarray}
D&=&\frac{3}{2}\left(\frac{m}{2l}\right)^3\,+\,{\cal O}(4),
\label{51}
\\
\t D&=&\frac{3}{2}\left(\frac{m}{2l}\right)^2\,+\,{\cal O}(3).
\label{52}
\end{eqnarray}
The power of 2 in (\ref{52}) seems in conflict with the 
usual ``tidal-force'' interpretation. The shapes themselves are also 
different. Like eggs with the thick ends pointing towards each 
other in the first case, and prolonged symmetrically (with respect 
to reflections on the equator $\eta=\pi/2$) in the second.

\end{document}